\documentclass[aps,prl,twocolumn]{revtex4}
\usepackage{amssymb,graphicx}
\usepackage{times}
\newlength{\bilderlength}

\newlength{\figsize}
\newcommand{\rme}{{\mathrm{e}}}
\newcommand{\rmd}{{\mathrm{d}}}

\begin{document}
\bibliographystyle{KAY}
\title{\sffamily \bfseries \Large
Stability of Random-Field and Random-Anisotropy Fixed Points at
large $N$}
\author{\sffamily\bfseries\normalsize Pierre Le Doussal and Kay J\"org Wiese \vspace*{3mm}} \affiliation{
CNRS-Laboratoire de Physique Th{\'e}orique de l'Ecole Normale
Sup{\'e}rieure, 24 rue Lhomond, 75231 Paris Cedex, France. }

\date{\small\today}

\begin{abstract}
In this note, we clarify the stability of the large-$N$ functional RG
fixed points of the order/disorder transition in the random-field (RF)
and random-anisotropy (RA) $O(N)$ models. We carefully distinguish
between infinite $N$, and large but finite $N$. For infinite $N$, the
Schwarz-Soffer inequality does not give a useful bound, and all fixed
points found in Phys. Rev. Lett. 96, 197202 (2006) (cond-mat/0510344)
correspond to physical disorder. For large but finite $N$ (i.e. to
first order in $1/N$) the non-analytic RF fixed point becomes
unstable, and the disorder flows to an analytic fixed point
characterized by dimensional reduction. However, for random anisotropy
the fixed point remains non-analytic (i.e.\ exhibits a cusp) and is
stable in the $1/N$ expansion, while the corresponding
dimensional-reduction fixed point is unstable. In this case the
Schwarz-Soffer inequality does not constrain the 2-point spin
correlation. We compute the critical exponents of this new fixed point
in a series in $1/N$ and to 2-loop order.
\end{abstract}
\maketitle

The random field (RF) and random anisotropy (RA) $N$-vector model is
studied by expanding around the 4-dimensional non-linear
$\sigma$-model \cite{Fisher1985b}. To this aim consider $O(N)$
classical spins $\vec{n}(x)$ with $N$ components and of unit norm
$\vec{n}^2=1$. To describe disorder-averaged correlations one
introduces replicas $\vec{n}_a(x)$, $a=1,\dots, k$, the limit $k=0$
being implicit everywhere. This gives a  non-linear sigma model, of
partition function ${\cal Z}=\int {\cal D} [\pi]\, \rme^{-{\cal
S}[\pi]}$ and action:
\begin{eqnarray}\label{action}
 {\cal S}[\pi ] = \int \rmd^d x &\Big[&\! \frac{1}{2 T_{0}} \sum_a [ (\nabla
\vec \pi_a)^2 + (\nabla \sigma_a)^2 ] - \frac{1}{T_{0}} \sum_a M_0 \sigma_a
\nonumber  \\
&& - \frac{1}{2 T_0^2} \sum_{a b} \hat R_0(\vec n_a \vec n_b) \Big]\ ,
\end{eqnarray}
where $\vec n_a=(\sigma_a,\vec \pi_a)$ with
$\sigma_a(x)=\sqrt{1-\vec \pi_a(x)^2}$. A small uniform external
field $\sim M_0 (1,\vec 0)$ acts as an infrared cutoff. The
ferromagnetic exchange produces the 1-replica part, while the random
field yields the 2-replica term $\hat R_0(z) = z$ for a bare
Gaussian RF. Random anisotropy corresponds to $\hat R_0(z) = z^2$.
As shown in \cite{Fisher1985b} one must include a full function
$\hat R_0(z)$, as it is generated under RG. It is marginal in $d=4$.

Recently, we have obtained results at 2-loop order \cite{us}, and
large $N$ for the ferromagnetic to disorder transition. In Ref.
\cite{comment} the authors argue that the large-$N$ fixed points
obtained by us (given after Eq.~(10) in \cite{us}) are unstable.
Here we reply to their argument.

The authors of Ref. \cite{comment} correctly point out that the
Schwartz-Soffer (SS) inequalities \cite{SS} put useful constraints
on the phase diagram of the {\it random-field} $O(N)$ model and its
(subtle) dependence in $N$. In our Letter \cite{us} we have studied
the Functional RG at large $N$ and obtained a series of fixed points
indexed by $n=2,3\dots $ where the disorder correlator $\hat R(z)$
(notations of \cite{us}) has a non-analyticity at $z=1$.  The $n=2$
fixed point (FP) has random field symmetry (RF) and $n=3$ has random
anisotropy (RA) symmetry ($\hat R(z)$ even in $z$). In addition we
found two infinite-$N$ analytic fixed points which obey dimensional
reduction. One of them ($\hat R(z)=z-1/2$) is the large-$N$ limit of
the Tarjus-Tissier (TT) FP \cite{tt} which exists for $N>N^*$ (at
two loop we found $N^*=18-\frac{49}{5} \bar\epsilon$,
$\bar\epsilon=d-4 \geq 0$) and has a weaker and weaker ``subcusp''
non-analyticity as $N$ increases. The question is which of these FPs
describes the ferromagnetic/disordered (FD) transition at large $N$
for $d\geq 4$.

First one should carefully distinguish: {\em (i)} strictly infinite
$N=\infty$ from large but finite $N$, 
{\em (ii)} RF symmetry vs.\ RA. We have shown \cite{us,us_long} that
for RF at $N=\infty$ physical initial conditions on the critical FD
manifold converge to the $n=2$ FP if the bare disorder is strong
enough ($r_4>4$ in \cite{us}). Hence for $N=\infty$ all these
non-analytic (NA) FPs are consistent. One can indeed check that they
correspond to a positive probability distribution of the disorder
since all $\hat R^{({n})}(0)$, the variances of the corresponding
random fields and anisotropies, are positive -- a condition hereby
referred to as physical. Furthermore the SS inequality does not yield
any useful constraint at $N=\infty$ because it contains an amplitude
itself proportional to $\sqrt{N}$.

Next, each of the above FPs can be followed down to finite $N$,
within an $1/N$ expansion performed to a high order in Ref.
\cite{us_long,us_florian}. It yields (to first order in
$\bar\epsilon=d-4$) the critical exponents $\bar \eta(n,N)$ and
$\eta(n,N)$ to high orders in $1/N$. One finds that the $n=2$ FP
acquires a {\it negative} $\hat R'(0)$ at order $1/N$, $\hat R'(0) =
-\frac{3}{4}\frac{\bar \epsilon}{N^{2}}+O (\frac{1}{N^{3}})$; hence
it becomes unphysical at finite $N$, a fact consistent with the
violation of the SS inequality $\bar \eta \leq 2 \eta$ correctly
pointed out in \cite{comment}. A natural scenario for RF symmetry,
as we indicated in our Letter \cite{us}, is that the FRG flows to
the TT FP for any {\it finite} $N>N^*$. However, as we discussed
there, if bare disorder is strong enough, it may approach the TT FP
along a NA direction, since these arguments relied only on blowing
up of $R''''(0)$ ($R(\phi)= \hat R(z=\cos(\phi))$).

A very interesting point, missed in Ref. \cite{comment}, is that the
SS inequalities do not constrain the 2-point function of the spin
$S^{i} (x)$ for {\it random anisotropy} disorder (it only constrains
the 2-point function of $\chi_{ij} (x)= S^{i} (x) S^{j} (x)$ as
disorder couples to the latter). Furthermore we find
\cite{us_long,us_florian} that the $n=3$ random anisotropy FP (which
reads $N {R(\phi)}/{|\epsilon| } = \frac{9}{8}\big ( 2 \cos(\phi)
\cos(\frac{\phi+\pi}{3}) + \cos(\frac{\pi-2 \phi}{3}) - 1 \big)$ in
the $N=\infty$ limit) {\it remains physical} for finite $N$.
Denoting $\hat R(z) = \bar\epsilon \mu \tilde R(z)$ with $\mu
=\frac{1}{N-2}$ and $y_0=\tilde R'(1)$, we obtain the following
expansion to $O(\bar \epsilon)$ for the exponents $\eta = y_{0}
\bar\epsilon/(N-2)$, $\bar \eta = (\frac{N-1}{N-2} y_{0}-1)
\bar\epsilon$, where

\begin{eqnarray}
y_0 &=&\frac{3}{2}+23 \mu -\frac{1750 \mu ^2}{3}+\frac{2129692 \mu^3}{27}-\frac{13386562376 \mu^4}{1215} \nonumber \\
&+&\! \frac{2004388412086052 \mu^5}{1148175}-\frac{107423933633514594598
\mu^6}{361675125}\nonumber \\
&+&\!\frac{66496428379374257425781597 \mu^7}{1253204308125}
+O\left(\mu ^9 \right)
\end{eqnarray}
and all coefficients in the expansion of $\hat{R}^{(n)}(0)$ near
$z=0$ remain indeed positive, e.g.:
\begin{eqnarray}\label{c49}
\tilde R' (z) = \Big[\frac{70 \mu }{9}{+}1 \Big] z+ \Big[\frac{1192
\mu }{243}{+}\frac{4}{27} \Big] z^3+ \Big[\frac{4384 \mu
}{2187}{+}\frac{16}{243}\Big] z^5  \nonumber \\
 + \Big[\frac{68608 \mu }{59049}{+}\frac{256}{6561} \Big] z^7 +
\Big[\frac{3735040 \mu }{4782969} {+}\frac{14080}{531441} \Big] z^9+O(z^{11})
\nonumber
\end{eqnarray}
Finally, for the $1/N$ expansion of the {\it
analytic} (DR) FP corresponding to RA we obtain (with $y_0=1$):
\begin{eqnarray}\label{c49}
\tilde R (z)&=&  \frac{z^2}{2}+\left(- \frac{3}{2} + 4 z^2-2
z^4\right) \mu+\dots\ ,
\end{eqnarray}
hence it becomes {\it unphysical} at finite $N$ \cite{footnote}. The
scenario is thus the opposite of the RF case: The NA FP $n=3$ is the
only one physical at large $N$ (it exists for $N>N_c=9.44121$) and
has precisely one unstable eigenvector (within the RA symmetry) as
expected for the FD transition. Using our 2-loop result \cite{us}
we  further obtained, up to $O\!\left(\mu ^2\right)$: $y_0=
\frac{3}{2} + 23 \mu +\left ( 9 \gamma _a-\frac{97}{4}\right) \mu \bar
\epsilon$, $\eta = \mu (\frac{3}{2}  \bar \epsilon + \bar \epsilon^2
(3 \gamma_a - \frac{27}{8}) )$ and $\bar \eta = \frac{\epsilon}{2} +
\mu (\frac{49}{2} \bar \epsilon + \bar \epsilon^2 (9 \gamma_a -
\frac{203}{8}) )$, where $\gamma_a$ was defined in \cite{us}.

Our conclusion is thus that the random anisotropy FP smoothly
matches to our solution $n=3$ at $N=\infty$ and remains
non-analytic for all $N$, breaking dimensional reduction. It does
not exhibit the TT phenomenon which seems a peculiarity of the RF
class. It is further studied in \cite{us_long,us_florian}.

We thank Florian K\"uhnel for useful discussions and acknowledge
support from ANR program 05-BLAN-0099-01. We also thank Gilles Tarjus
for pointing out that similar results were independently obtained in
\cite{TT2}.

\end{document}